\documentclass{Interspeech2024}
\usepackage{caption}
\usepackage{subcaption}
\usepackage{dsfont}
\usepackage{eucal}
\usepackage{arydshln}
\usepackage{relsize}
\usepackage{threeparttable}

\title{An efficient text augmentation approach for contextualized Mandarin speech recognition}

\name{Naijun}{Zheng}
\name{Xucheng}{Wan}
\name{Kai}{Liu}
\name{Ziqing}{Du}
\name{Zhou}{Huan}

\address{IT Innovation and Research Center, Huawei Technologies, China}
\email{\{zhengnaijun, zhou.huan\}@huawei.com}


\begin{document}

\maketitle

\begin{abstract}
    Although contextualized automatic speech recognition (ASR) systems are commonly used to improve the recognition of uncommon words, their effectiveness is hindered by the inherent limitations of speech-text data availability. To address this challenge, our study proposes to leverage extensive text-only datasets and contextualize pre-trained ASR models using a straightforward text-augmentation (TA) technique, all while keeping computational costs minimal. In particular, to contextualize a pre-trained CIF-based ASR, we construct a codebook using limited speech-text data. By utilizing a simple codebook lookup process, we convert available text-only data into latent text embeddings. These embeddings then enhance the inputs for the contextualized ASR. Our experiments on diverse Mandarin test sets demonstrate that our TA approach significantly boosts recognition performance. The top-performing system shows relative CER improvements of up to 30\% on rare words and 15\% across all words in general.
    
\end{abstract}

\section{Introduction}

To transcribe human speech into text, End-to-end (E2E) automatic speech recognition (ASR) has been widely studied. Through supervised training on speech-text pairs, various E2E ASR architectures, such as connectionist temporal classification (CTC) \cite{graves_ctc}, recurrent neural network transducer (RNN-T) \cite{pmlr-v32-graves14}, attention-based encoder-decoder (AED) \cite{Chorowski_15attention, dong_18transformer, gulati20_interspeech} have achieved impressive results. Nevertheless, these E2E ASR systems might encounter recognition errors with long-tailed rare words (such as jargon or unusual named-entities in unique target domain), presenting a challenge for real-world ASR implementations..

To address this issue, the predominant focus of research is on enhancing an E2E ASR system by integrating contextual information (extracted from a predefined list of rare words in the target domain), known as contextualized ASR \cite{kan_18_sf, zhao19d_interspeech, jain20_interspeech}.
Among them, a popular way is deep contextualization, where neural modules (based on trie approaches \cite{le21_interspeech, sun_23_trie} or attention approaches \cite{pundak_18_clas, xu23d_interspeech, cif_han2021, huang23_spike, shi2023seacoparaformer}) are integrated to learn contextual information through supervised training using speech-text pairs. 
Recently, deep contextualization approaches with explicit contextual decoding and bias loss demonstrate promising results in CTC-based \cite{huang23_spike}, transducer-based ASR \cite{wang2023_spell2} and AED-based ASR employing the continuous integrate-and-fire (CIF) mechanism \cite{cif_han2021, cif2_han2022}.
In practical terms, the limited availability of paired speech-text data reduces the effectiveness of deep contextualization methods. In contrast, unpaired text data specific to the target domain are more abundant, easier to obtain and have a richer vocabulary coverage. This makes unpaired text data as a good candidate for learning contextual information.

However, little prior research is found in the aforementioned direction. One related work is \cite{qiu23_ustr}, which leverages USTR \cite{huang23f_interspeech} to incorporate text data using a shared encoder to enhance the biasing module, then conduct fine-tuning on an RNN-T based ASR. The same concept has been explored by many studies focusing on address typical domain mismatch problems in ASR \cite{sainath_joist, chen22r_interspeech, zhang2022speechlm, chen_23_ctctext}. Most approaches employ a speech encoder and a text encoder to map the speech and text input to a unified space via joint optimization.
In conclusion, all of these methods have a limitation in that fine-tuning the ASR for the target domain is required, which may not always be feasible, primarily due to computational constraints. 

Given practical resource constraints concerning both paired data and computational resources, the aim of this study is to investigate an economical and practical contextual ASR approach that fully utilizes additional unpaired text data. In essence its benefits include: 1) eliminating the need for ASR fine-tuning; 2) requiring only a small amount of paired data in the target domain; and 3) enhancing recognition of rare words by utilizing extensive unpaired text data. To the best of our knowledge, this study could be the initial endeavor to create a contextualized ASR under the aforementioned constraints.
During text data learning, the alignment information between speech and text is not accessible. Therefore, in this paper, we depend on a CIF-based ASR. Without the need for any additional features, we introduce a codebook sampler to implicitly align sequences at an extremely low cost. We also investigate multiple TA approaches for biasing coding using various setups of the codebook samplers. 
As we will demonstrate, experiments conducted on various test sets indicate that all corresponding TA-enhanced contextualized ASR systems significantly outperform the baseline system.
In the rest of the paper, we first review the ASR backbone with the CIF mechanism in Section 2, and describe our contextual ASR with text augmentation in Section 3. Experimental results are presented in Section 4 and Section 5 concludes the paper.

\section{Preliminaries}
\label{sec:asr}


\textbf{Continuous integrate-and-fire (CIF)} \cite{cif_dong2020} Inspired by the integrate-and-fire scheme, CIF was proposed for an encoder-decoder-based ASR framework. It integrates frame-level acoustic embedding sequences and sends the integrated embeddings to the decoder for token prediction. As shown in Fig.\ref{fig:system}(a), given an utterance $u$ with frame-level feature $X_{1:T}$ and label $Y_{1:N_\text{ref}}$ (where $N$ is the number of tokens), CIF accumulates encoded input $\{e_{1:T}\}$ according to the estimated weights for each frame and outputs the integrated embeddings ${E}_{\text{cif}}\in \mathcal{R}^{D*N_\text{hyp}}$, i.e., $[N_{\text{hyp}}, {E}_{\text{cif}}] = \text{CIF}(e_{1:T})$. 
During training, the initial $N_\text{hyp}$ is not expected to be the same as $N_\text{ref}$. When it occurs, a scaling strategy is applied by teacher-forcing CIF to re-produce ${E}_\text{cif}\in \mathcal{R}^{D*N_\text{ref}}$. 

\noindent\textbf{Contextualized ASR} \cite{cif_han2021, cif2_han2022,shi2023seacoparaformer}
To enhance the recognition capability of user-defined list of rare phrases (also know as hotwords), several contextualized ASR approaches have been proposed to supplement a basic E2E ASR with a separate biasing module. The bias module typically comprises two components: a bias encoder and an attention-based bias decoder. The encoder's goal is to transform the hotword list $Y^h$ (with $N_h$ phrases) into word-level embeddings $Z$, that is, $Z_{1:N_h} = \text{BiasEnc}(Y^h_{1:N_h})$. In addition, an extra \textit{no-bias} option is appended to the hotword list to manage the case without bias.
While the bias decoder takes inputs from both latent speech feature $E_{\text{speech}}$ and hotword embeddings \(Z\). With application of an attention mechanism, the bias decoder explicitly predicts the hotword probability, $P_{1:N_\text{hyp}} = \text{BiasDec}(E_{\text{speech},1:N_\text{hyp}}, Z_{1:N_h+1})$.
Optionally, $E_{\text{speech}}$ can be modeled either by the feature space $\mathcal{S}_{cif}$ from CIF output or by the feature space $\mathcal{S}_{dec}$ from an intermediate decoder output (e.g., the last layer in the decoder \cite{shi2023seacoparaformer}).
During inference, once occurrence of a particular hotword is predicted, the ASR hypothesis is combined with the hotword through collaborative decoding.




\section{Proposed Method}
\label{sec: method}

\subsection{Framework pipeline}
\begin{figure}
    \centering
    \includegraphics[width=9.0cm]{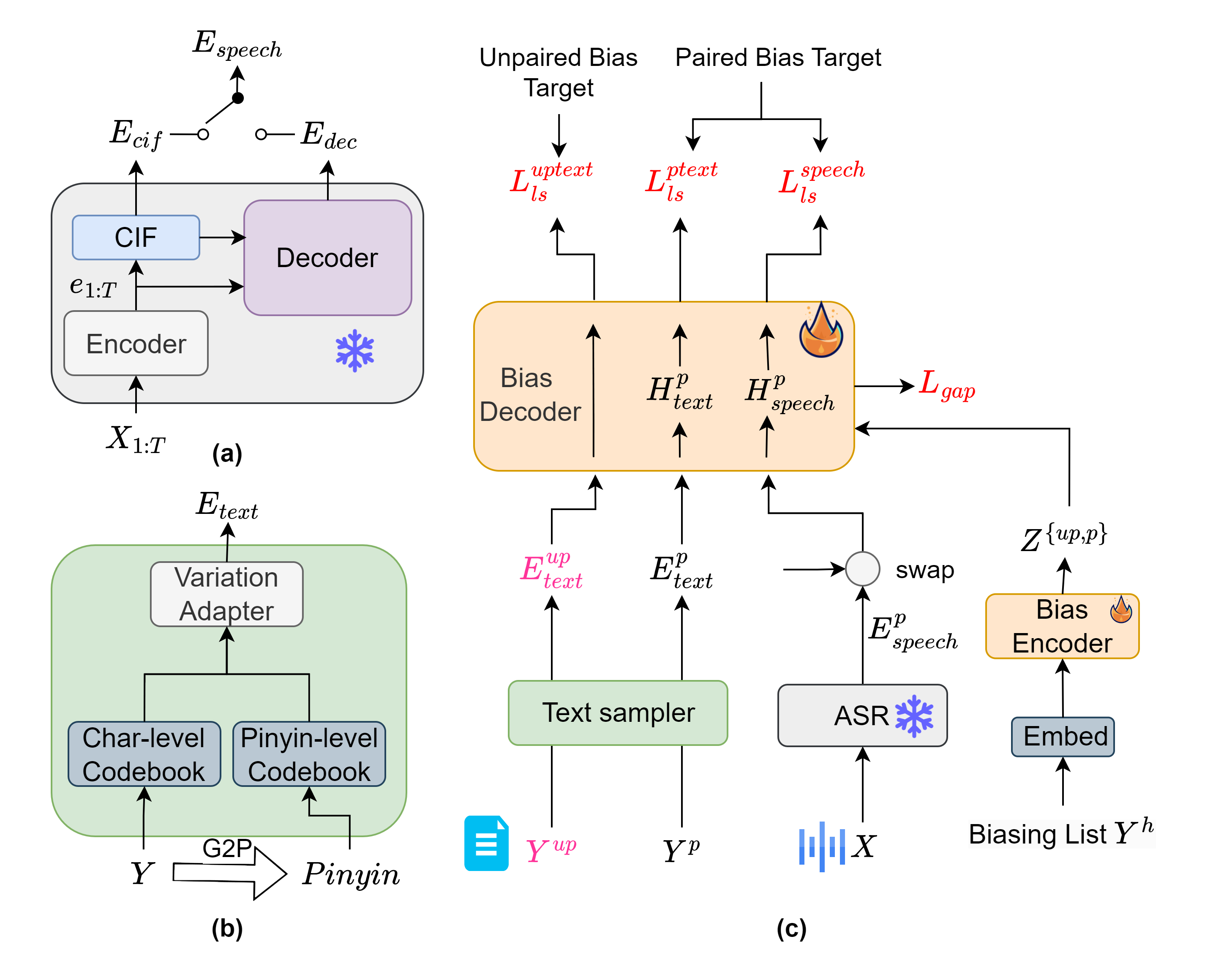}
    \caption{Overall diagram of proposed TA-enhanced contextualized ASR system: (a) CIF-based ASR; (b) Text sampler; (c) Training framework with TA-enhanced biasing module.}
    \label{fig:system}
\end{figure}

Fig.\ref{fig:system}(c) illustrates the framework of our proposed TA-enhanced contextualized ASR. It consists of three components: CIF-based ASR module, text sampler module and biasing module (with encoder and decoder). In the context, the ASR module refers to a CIF-based network with fixed parameters acquired from pre-training in the source domain. 
To enrich input for the biasing module, the implementation pipeline of our TA approach comprises three steps: Firstly, a text sampler module is developed using paired speech-text data. Next, the available text data (including both paired transcriptions and unpaired text) is embedded into the hidden latent space $\mathcal{S}$. Finally, the combined embeddings of speech and text are inputted into the bias decoder during training.

\subsection{Text sampler module}
The text sampler module aims to transform textual input $Y$ into hidden embeddings $E_\text{text}\in\mathcal{S}$. Since modality matching between speech and text can be only performed on paired data, we propose to use a small amount of paired data to initially construct a codebook (CB), with leveraging a pre-trained CIF-based ASR.

\noindent\textbf{CB construction}
Given a piece of paired speech training data with speech feature $X^i$ and transcription label $Y^i$, its ASR embedding $E^i_{\text{speech}} \in \mathcal{S}$ can be extracted.
Considering the case that ${N}^i_{\text{hyp}}\neq N^i_{\text{ref}}$, a scaling strategy is applied for length calibration. Specially, we introduce a confidence index $w^i \in [0,1]$, defined by $w^i = 1-\min\left({{\vert N^i_{hyp}-N^i_{ref}\vert}/{N^i_{ref}}, 1}\right)$. That is, a triplet output is generated from the ASR module, 
$[N^i_{\text{ref}}, w^i, {E}^i_{\text{speech}}] = \text{ASR}(X^i, Y^i)$.

By performing the process on all $N_p$ paired data, each text token presented in training transcriptions may end up with multiple embedding representations. By assuming that these embeddings follow a Gaussian distribution, the codebook embedding for a given token $c$ can be computed as the center vector: 
\begin{equation}
    \text{CB}_\text{token}(c) = \frac{\sum_{i=1}^{N_p}w^i\sum_{j}E^i_{\text{speech},j}\delta(u_j=c)}{\sum_{i=1}^{N_p}w^i\sum_{j}\delta(u_j=c)}, 
    \label{eq:embtable}
\end{equation}
where $E^i_{\text{speech},j}$ represents the speech embedding unit aligned to the $j$-th token in the $i$-th utterance, $\delta$ denotes the Dirac delta function.

\noindent\textbf{CB configuration} To investigate the effectiveness of space $\mathcal{S}$, we explore two alternative latent spaces: $\mathcal{S}_\text{cif}$ from the ASR encoder (more acoustic-like) and $\mathcal{S}_\text{dec}$ from the decoder (more linguistic like), respectively. In addition, for Mandarin, text units can be modeled by either Chinese characters (representing meaning) or Pinyin\footnote{A system used to represent the pronunciation of Chinese characters using the Latin alphabet, including syllables and tones.} syllables (relating to sound). As such, two CBs are built to accommodate the two different modeling units. To sum up, total four CBs are configured, $\text{CB}_\text{char,cif}, \text{CB}_\text{pinyin,cif}$ that represent character and Pinyin in latent space $\mathcal{S}_\text{cif}$ and $\text{CB}_\text{char,dec}, \text{CB}_\text{pinyin,dec}$ in space $\mathcal{S}_\text{dec}$, respectively. 

\begin{figure}
    \centering
    \begin{subfigure}[b]{0.49\linewidth}
        \centering
        \includegraphics[width=\linewidth]{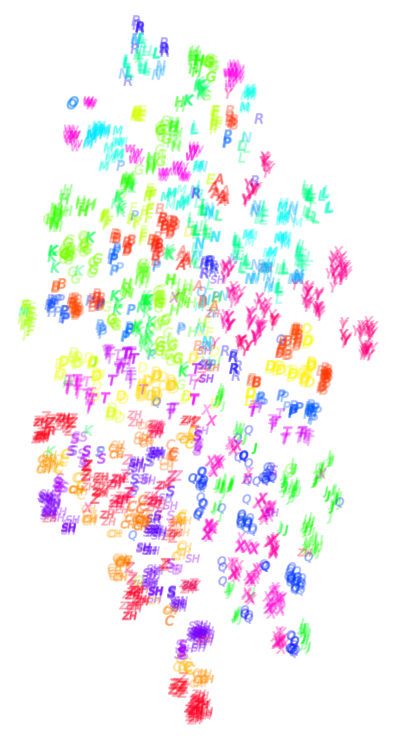}
        \caption{CIF space.}
        \label{fig:cif_emb}
    \end{subfigure}
    \hfill
    \begin{subfigure}[b]{0.49\linewidth}
        \centering
        \includegraphics[width=\linewidth]{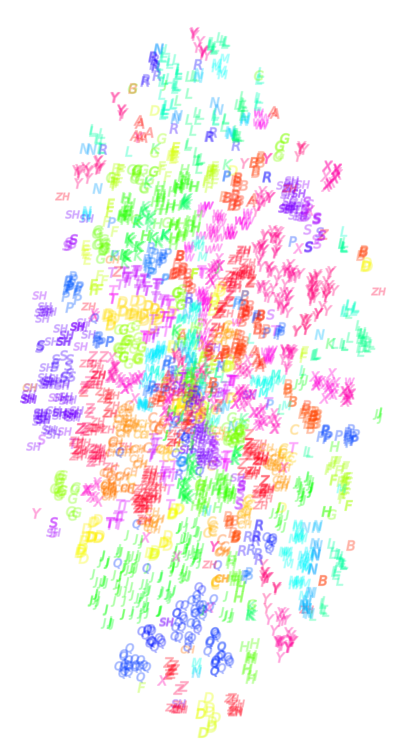}
        \caption{Decoder space.}
        \label{fig:dec_emb}
    \end{subfigure}
    \caption{Distribution plots for text embeddings of Chinese characters, where each point is labeled and colored according to the first syllable in Pinyin. (Best viewed in color)}
    \vspace{-0.5cm}
    \label{fig:emb_dis}
\end{figure}

\noindent\textbf{CB-lookup} As a preliminary study, Fig.\ref{fig:emb_dis} shows the embedding distributions from $\text{CB}_\text{char,cif}$ (left sub-figure) and $\text{CB}_\text{char,dec}$ in (right sub-figure), visualized via the t-SNE technique \cite{vandermaaten08a}. As expected, embeddings with similar pronunciations are clustered together, and clusters in (a) are more compact than those in (b), showing smaller within-cluster variances. Following this observation, we can easily transform unpaired text $Y^{up}$ into hidden embedding $E^{up}_\text{text}$ and paired transcripts $Y^p$ into $E^p_\text{text}$ subsequently, via a simple CB lookup.
Note that since unpaired data has a larger scale, some words may be unseen during the construction of the CB, which may result in no direct lookup results in the CB. For these out-of-distribution words, we first utilize a Chinese tokenizer to segment the utterance text $u$ and deploy $\text{CB}_\text{pinyin}$ with a grapheme to phoneme (G2P) to map the token $c$ into an embedding.  
\begin{equation*}
    E_\text{text}(c, u) = \biggl\{
    \begin{aligned}
        &\text{CB}_\text{char}(c), \text{if $c$ in CB$_\text{char}$}  \\ &\text{CB}_\text{pinyin}(\textit{G2P}(\text{Tokenizer}(c,u))), \text{otherwise}
    \end{aligned}
\end{equation*}

\subsection{Optional Variation Adaption (VA) schemes}
\label{ssec:va}
As shown in Eq.\eqref{eq:embtable}, $\text{CB}_{token}$ of each character or Pinyin can be straightforwardly obtained. However, in practice, such embeddings are expected to be somewhat flexible, deviating from the central representations due to the speaker's accent, emotion, contextual texts, or other forms of variability.
To address the real concern, we deploy an optional variation adaptor on top of the CBs (referring to Fig.\ref{fig:system} (b)) to generate more flexible embeddings. 
Specifically, three alternative adaptation schemes are explored : 
(I) Use $\text{CB}_\text{pinyin}$ exclusively; 
(II) Replace the center vector in Eq.\eqref{eq:embtable} with a random sample from Gaussian distributions. 
(III) Randomly replace the characters with the other homophone characters that share the same pronunciation to simulate the scenario where ASR outputs are erroneous.

\subsection{Training and inference}
To train the biasing module, we  can employ two possible text sources, $Y^p$ and $Y^{up}$. Depending on the availability of the source, the query inputs for the biasing decoder may contain: oracle input $E_\text{speech}$, augmented input with $E^p_\text{text}$, and augmented input with $E^{up}_\text{text}$. As an exemplary figure, Fig.\ref{fig:system}(c) depicts an full TA approach with three types of combined inputs for the biasing decoder. 
A swapping mechanism \cite{zhang2022speechlm} is also applied by randomly replacing some elements in $E_\text{speech}^p$ with $E_\text{text}^p$.


In addition, to explicitly bridge the modality gap between speech and text, we create an auxiliary loss function to supplement the regular biasing loss. Based on the hidden representations $H_\text{speech}$ and $H_\text{text}$ (extracted from an intermediate bias decoder layer), we apply a cosine-distance-based loss as penalty to reduce the similarity gap,  calculated by:
\begin{equation}
    L_{gap} = \frac{1}{\sum_{i=0}^{B-1}w^i}\sum_{u=0}^{B-1}w^i(1 - \cos(H^p_{speech}, H^p_{text})),
    \label{eq:cos}
\end{equation}
where $B$ denotes the batch size, $w$ is the same confidence index used for CB construction. In turn, our overall objective becomes:
\begin{equation}
    L = L_{ls}^{speech} + \lambda(L_{ls}^{p_{text}} + L_{ls}^{up_{text}}) + L_{gap},
    \label{eq:loss}
\end{equation}
where $L_{ls}$ denotes the label smoothing loss function \cite{chorowski2016better} for hotword prediction task using masked $Y$ as bias targets, and $\lambda$ is a hyper-parameter to control the penalty.

\noindent\textbf{Inference}
In inference, the contextual ASR performs regularly by discarding the text sampler module and loading $E_\text{speech}$ only to the bias decoder.
When bias decoded character with highest probability is not \textit{no-bias}, the contextual output replaces its counterpart presented in the ASR hypothesis. Meanwhile, to further enhance the performance, we also adopt the attention score filtering (ASF) \cite{shi2023seacoparaformer} technique, which picks the most active $k$ hotwords based on the cross attention scores and mask others during a second-time contextual bias decoding.


\section{Experiments}

\subsection{Data sets}

Our experiments are conducted on open-source Mandarin speech datasets, \textit{WeNetSpeech} \cite{zhang2022wenetspeech} and \textit{Aishell1} \cite{aishell_2017}. Data details are shown in Table \ref{tab:dataset}. Note that we chose the transcripts from \textit{Wenetspeech large} as text-only dataset (no overlap with test sets), which is nine times larger than the paired data. 

Four test sets are used for performance evaluation. Considering the \textit{Test-bias} \cite{Xu_cb2023} (including 298 hotwords) is less challenging for a well-trained ASR, we created a new test-set from the \textit{WenetSpeech test-net set} using the Hanlp tool \cite{he-choi-2021-stem} to identify the named entities (for person, organization and place).
In addition, we also adopt two \textit{Aishell}-related test sets as out-of-domain datasets. Like prior works, We use two metrics:  Character Error Rate (CER) for general text and Biased CER (B-CER) \cite{le21_interspeech,huang23_spike} for characters in the hotword list.

\begin{table}[htbp]
  \centering
  \caption{Information of data sets}
    \scalebox{0.7}{
    \begin{tabular}{rllcc}
    \hline
          & Name  & Source & \#utt & \#hotword \\
    \hline
    \multicolumn{1}{l}{Train} & Paried speech & Wenetspeech middle & 1,277k & - \\
          & Unparied text & Wenetspeech large & 10,144k & - \\
    \hline
    \multicolumn{1}{l}{Dev} & Dev-bias & Wenetspeech middle & 2,497  & - \\
    \hline
    \multicolumn{1}{l}{Test} & Test-bias & Wenetspeech middle & 21,972 & 298 \\
          & Test-net bias & Wenetspeech test\_net & 5,311  & 1,523 \\
          & Dev-Aishell1-NE  & Aishell1 dev set & 1,334  & 600 \\
          & Test-Aishell1-NE & Aishell1 test set & 808   & 400 \\
    \hline
    \hline
    \end{tabular}%
    }
  \label{tab:dataset}%
\end{table}%

\begin{table*}[htbp]
  \centering
  \caption{CER Results over all test sets, where TA denotes the proposed text augmentation.}
\scalebox{0.8}{
\begin{threeparttable}
\begin{tabular}{l|cc|cc|cc|cc}
\hline
      & \multicolumn{4}{c|}{In-domain} & \multicolumn{4}{c}{Out-of-domain} \\
\cline{2-9}Systems & \multicolumn{2}{c|}{Test-bias (k=50)} & \multicolumn{2}{c|}{Test-net bias (k=50)} & \multicolumn{2}{c|}{ Dev-Aishell NE (k=10)} & \multicolumn{2}{c}{Test-Aishell NE (k=10)} \\
\cline{2-9}      & CER   & B-CER & CER   & B-CER & CER   & B-CER & CER   & B-CER \\
\hline
B0 (raw ASR) & 7.21  & 8.83  & 9.60  & 25.86  & 8.11  & 24.73  & 8.47  & 23.57  \\
\hline
B1 (ASR + CIF bias) & 7.09  & 4.88  & 8.84  & 16.94  & 5.25  & 8.92  & 6.71  & 13.75  \\
B1+TA & \textbf{6.76} & 4.62  & 8.58  & 17.94  & 4.72  & 8.21  & 5.03  & 7.80  \\
\hdashline
B1*   & 6.97  & 4.52  & 8.74  & 15.37  & 4.97  & 8.56  & 6.16  & 12.07  \\
B1+TA* & 6.84  & \textbf{4.47} & \textbf{8.29} & \textbf{13.82} & \textbf{4.21} & \textbf{4.89} & \textbf{4.50} & \textbf{4.48} \\
\hline
B2 (ASR + DEC bias) & 6.56  & 4.85  & 9.13  & 22.32  & 7.21  & 20.08  & 8.05  & 21.30  \\
B2+TA & 6.50  & 4.41  & 9.09  & 22.02  & 6.36  & 16.39  & 6.74  & 15.29  \\
\hdashline
B2*   & 6.47  & 4.18  & 8.74  & 19.20  & 6.51  & 16.46  & 7.22  & 17.21  \\
B2+TA* & \textbf{6.45} & \textbf{3.98} & \textbf{8.53} & \textbf{17.81} & \textbf{5.07} & \textbf{10.16} & \textbf{5.58} & \textbf{9.83} \\
\hline
B3 (ASR + CIF-DEC bias) & 6.51  & 3.98  & 9.23  & 21.70  & 7.20  & 18.15  & 7.58  & 17.98  \\
B3+TA & 6.45  & 3.90  & 8.74  & 19.35  & 5.47  & 12.19  & 5.88  & 11.40  \\
\hdashline
B3*   & 6.46  & 3.74  & 8.61  & 17.69  & 5.70  & 12.21  & 6.45  & 13.64  \\
B3+TA* & \textbf{6.43} & \textbf{3.54} & \textbf{8.23} & \textbf{15.09} & \textbf{4.86} & \textbf{8.63} & \textbf{5.03} & \textbf{7.52} \\
\hline
\hline
\end{tabular}%

\begin{tablenotes}
       \footnotesize
       \item[1] {\small The symbol * denotes the ASF scheme is applied with k most active hotwords.}
       \item[2] {\small System B3 utilizes joint CIF-DEC embeddings, which combine CIF- and DEC-embeddings through element-wise summarization in the linear layer of the bias decoder.}
     \end{tablenotes}
    \end{threeparttable}

  \label{tab:cb}%
}
\end{table*}%

\subsection{Experimental setup}
The pre-trained Paraformer model\footnote{\small \url{http://modelscope.cn/models/iic/speech_paraformer_asr_nat-zh-cn-16k-common-vocab8358-tensorflow1}} \cite{gao22b_interspeech}  is used as our ASR backbone since it currently achieves super performance on most public Mandarin speech corpora.
Our bias encoder is trained from scratch. It consists of two BiLSTM layers in the encoder, four Transformer decoder layers and a linear output layer in the decoder. The hidden representations $H$ in Eq.\eqref{eq:cos} are extracted from the second bias decoder layer, and $\lambda$ in Eq.\eqref{eq:loss} is set to 0.1.
With learning rate of 0.0005 (utilizing warmup for the initial 3 steps), our model was trained for 50 epochs, and the final model was created by averaging parameters from the top 5 models. For each training batch, $Y^h$ is constructed using the n-gram method ($2\le n \le 6$) from each utterance.
The Pinyin is obtained by a tokenizer\footnote{\url{http://github.com/fxsjy/jieba}} for text segmentation, followed by a Chinese G2P toolkit\footnote{\url{http://github.com/mozillazg/python-pinyin}}.
The same batch size is adopted for the paired and unpaired text inputs, while respective biasing lists are independently constructed.
The codebook created for Pinyin consists of 1,238 items, whereas the codebook for characters comprises 5,438 items. 

\subsection{Experimental results}
To validate the effectiveness and explore the best configuration for our TA-enhanced contextual ASR, we conducted extensive experiments on both in-domain and out-of-domain test sets with various TA setups. The results are reported in Table \ref{tab:cb}. The optimal performance for each TA configuration is emphasized in bold for every dataset. 
From the table, we can obtain a few valuable insights: 1) The oracle ASR acts as a strong benchmark, showing decent CERs across all test sets. Nonetheless, as anticipated, the B-CER diminishes, particularly on challenging datasets;
2) The contextual ASR, coupled with an independent biasing module, establishes robust baseline systems (B1/B2/B3) for hotword recognition (referring to performance improvements from B1/B2/B3 over B0);
3) All findings demonstrate a consistent trend: the proposed TA approach consistently improves CER and B-CER, showcasing significant enhancements across all conditions;
4) The TA-enhanced B3 and B1 emerge as the most effective on the in-domain and out-of-domain test sets respectively.
On the out-of-domain test sets, there's a relative improvement over 30\% in B-CER and 15\% in CER compared to system B0.
Note that due to space limitations, we report only the results from $\text{CB}_\text{char}$, while similar observations were made for $\text{CB}_\text{pinyin}$.



\begin{figure}
    \centering
    \includegraphics[width=8cm]{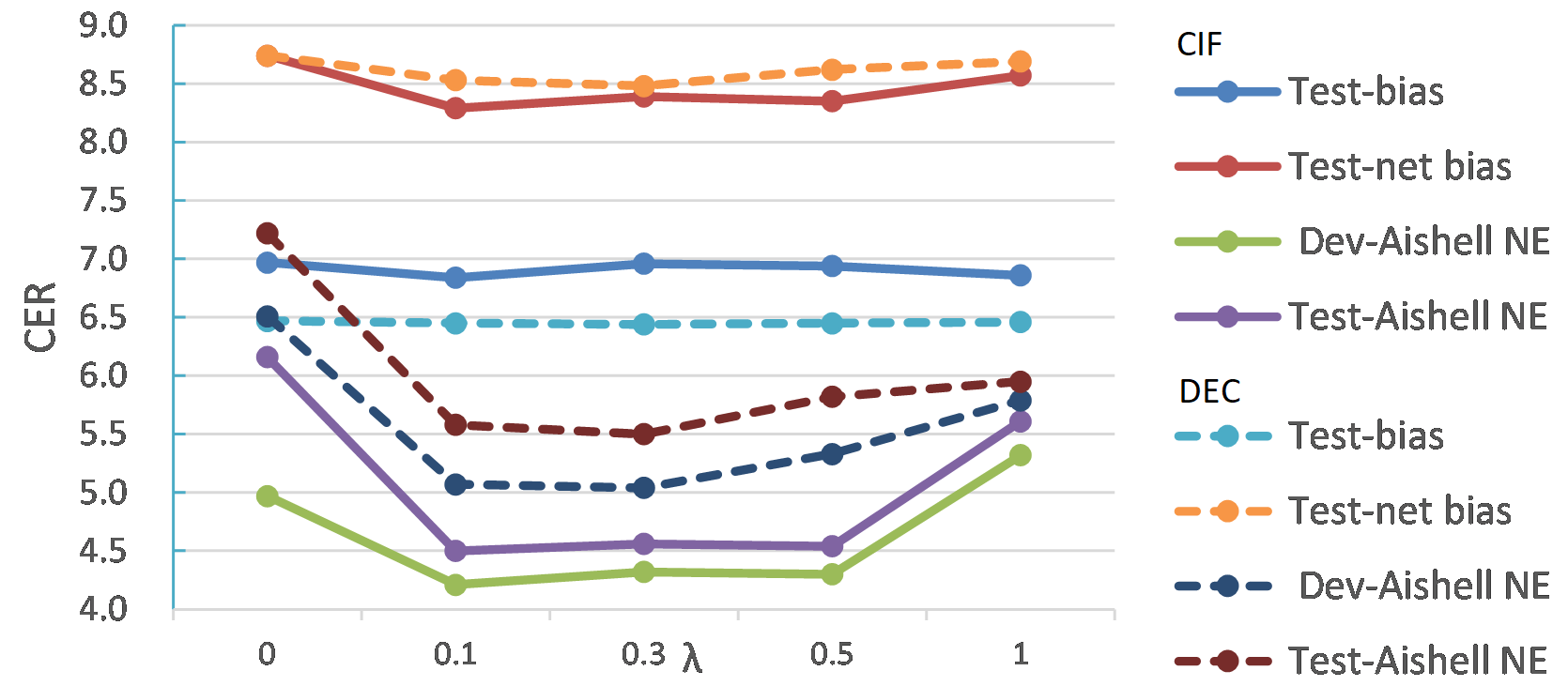}
    \caption{The impact of $\lambda$ on CER.}
    \label{fig:lambda}
\end{figure}

We also examine the impact of varying the parameter $\lambda$ in the training objective Eq.\eqref{eq:loss}. The results are shown in Fig.\ref{fig:lambda}. As expected, a small value for $\lambda$ (less than 0.5) is favored, as a larger value may lead to overfitting.

\begin{figure}
    \centering
    \includegraphics[width=7.8cm]{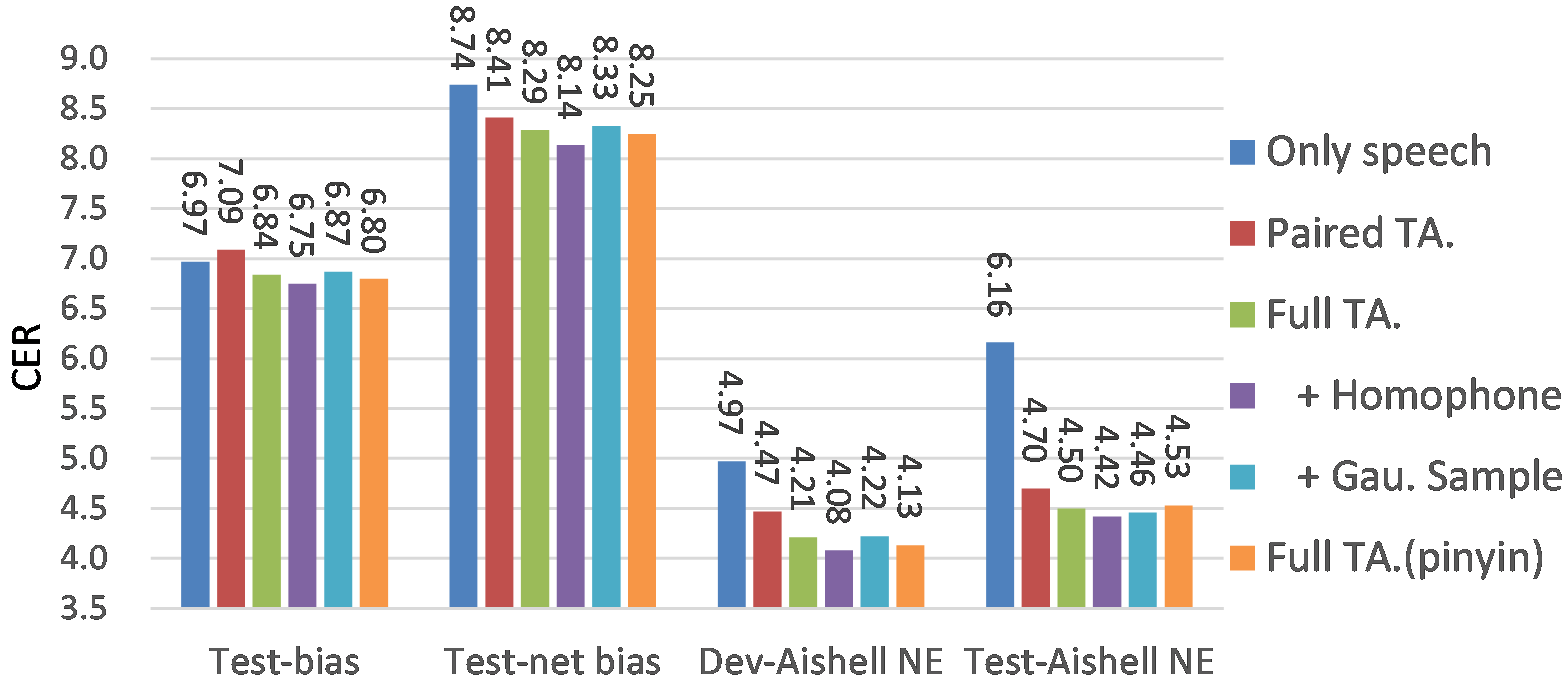}
    \caption{Comparisons of various VA schemes with CIF bias.}
    \label{fig:ablation}
\end{figure}

Finally, we investigate the impact of VA schemes based on system B1*. Moreover, to evaluate the system's capability to handle diverse inputs, we also explore a partial TA approach that utilizes only paired text for augmentation. Corresponding results are outlined in Fig.\ref{fig:ablation}. 
From the diagram, it's evident that full TA surpasses partial TA (using $Y^P_\text{text}$ only) in performance, owing to the presence of unpaired text (even with just 10k utterances). The last takeaway is about the effects of three proposed three VA schemes. We notice that the scheme I (exclusively using $\text{CB}_\text{pinyin}$) results in only marginal improvement in CER; contributions from the Gaussian sampling (scheme II) are negligible; whereas the third scheme (the homophone replacement with a ratio of 10\%) consistently enhances CER across all conditions. This reinforces our assertion that flexible text embeddings are preferred. 

\section{Conclusion}
In this study, to enhance ASR's ability in recognizing rare words, we introduced a straightforward yet effective text-augmented contextual biasing method for CIF-based ASR.
Our proposed approach presents several advantages for practical scenarios: it solely relies on text for augmentation; offers text embeddings at a minimal computational expense; removes the need for costly ASR re-training or excessively fine-tuning, and practically delivers significant performance improvements, even with very little text data (10k utterances). Moving forward, we intend to explore the application of our method in handling coding-switched hotwords. Additionally, we aim to expand our method to incorporate the TA approach with other E2E backbones.


\bibliographystyle{IEEEtran}
\bibliography{mybib}

\end{document}